\documentclass{aa}
\usepackage{graphicx}
\usepackage{natbib}
\usepackage{amstext}
\bibpunct{(}{)}{;}{a}{}{,}
\usepackage{txfonts} 
\usepackage{color}

\def\HII{\mbox{\ion{H}{ii}}}

\def\mag{\ifmmode^{\rm m }\else$^{\rm m}$\fi}

\def\as{$\,^{\prime\prime}\,$}

\def\hh{\ifmmode^{\rm h}\else$^{\rm h}$\fi}
\def\mm{\ifmmode^{\rm m}\else$^{\rm m}$\fi}
\def\ss{\ifmmode^{\rm s}\else$^{\rm s}$\fi}
\def\deg{\ifmmode^\circ\else$^\circ $\fi}
\def\amin{\ifmmode^\prime\else$^\prime $\fi}

\def\decdm#1#2{\ifmmode{#1}\else{$#1$}\fi\deg\ #2\amin\ }

\def\dec#1#2#3{\ifmmode{#1}\else{$#1$}\fi\deg\ #2\amin\ #3\as\ }

\def\decb#1#2#3#4{\ifmmode{#1}\else{$#1$}\fi\deg\ #2\amin\ #3\farcs#4 }

\def\torib{$\theta^1$Ori\,B}
\def\toric{$\theta^1$Ori\,C}
\def\torid{$\theta^1$Ori\,D}

\begin{document}

\title{Visual/infrared interferometry of Orion Trapezium stars: Preliminary
  dynamical orbit and aperture synthesis imaging of the $\theta^1$Orionis\,C
  system}

\titlerunning{Visual/infrared interferometry of Orion Trapezium stars}

\author{
   S.~Kraus\inst{1} \and 
   Y.~Y.~Balega\inst{2} \and
   J.-P.~Berger\inst{3} \and
   K.-H.~Hofmann\inst{1} \and
   R.~Millan-Gabet\inst{4} \and
   J.~D.~Monnier\inst{5} \and
   K.~Ohnaka\inst{1} \and
   E.~Pedretti\inst{5} \and
   Th.~Preibisch\inst{1} \and 
   D.~Schertl\inst{1} \and
   F.~P.~Schloerb\inst{6} \and
   W.~A.~Traub\inst{7} \and
   G.~Weigelt\inst{1}
}
 
 \offprints{skraus@mpifr-bonn.mpg.de}
 
 \institute{
  Max-Planck-Institut f\"ur Radioastronomie, Auf dem H\"ugel 69,
  D-53121 Bonn, Germany
  \and
  Special Astrophysical Observatory, Russian Academy of Sciences, Nizhnij
  Arkhyz, Zelenchuk region, Karachai-Cherkesia, 357147, Russia
  \and
  Laboratoire d'Astrophysique de Grenoble, UMR 5571 Universit\'e Joseph
  Fourier/CNRS, BP 53, F-38041 Grenoble Cedex 9, France
  \and
  Michelson Science Center, California Institute of Technology, Pasadena, CA
  91125, USA
  \and
  Astronomy Department, University of Michigan, 500 Church Street, Ann Arbor,
  MI 48104, USA
  \and
  Department of Astronomy, University of Massachusetts, LGRT-B 619E, 710 North
  Pleasant Street, Amherst, MA 01003, USA
  \and
  Harvard-Smithsonian Center for Astrophysics, 60 Garden Street, Cambridge, MA
  02183, USA
 }
 
 \date{Received December 19, 2006; accepted February 7, 2007}
 
  \abstract
  {Located in the Orion~Trapezium~cluster, {\toric} is one of the youngest
   and nearest high-mass stars (O5-O7) and known to be a close binary.
 }
 {By tracing its orbital motion, we aim to determine the orbit and
   dynamical mass of the system, yielding a characterization of the individual
   components and, ultimately, also new constraints for stellar evolution
   models in the high-mass regime.
 }
 {Using new multi-epoch visual and near-infrared bispectrum speckle
   interferometric observations obtained at the BTA~6~m~telescope, and
   IOTA near-infrared long-baseline interferometry, we traced the orbital
   motion of the {\toric} components over the interval 1997.8 to 2005.9,
   covering a significant arc of the orbit. Besides fitting the relative
   position and the flux ratio, we applied aperture synthesis techniques to our
   IOTA data to reconstruct a model-independent image of the {\toric} binary
   system. }
 {The orbital solutions suggest a highly eccentricity ($e\approx0.91$) and
   short-period ($P\approx10.9$~yrs) orbit.  As the current astrometric 
   data only allows rather weak constraints on the total dynamical mass, we
   present the two best-fit orbits.  Of these two, the one implying a system
   mass of $48~M_{\sun}$ and a distance of 434~pc to the Trapezium cluster can
   be favored.
   When also taking the measured flux ratio and the derived location in the
   HR-diagram into account, we find good agreement for all observables,
   assuming a spectral type of O5.5 for {\toric}1 
   ($M=34.0~M_{\sun}$,~$T_{\rm eff}=39\,900$~K) and O9.5 for C2
   ($M=15.5~M_{\sun}$,~$T_{\rm eff}=31\,900$~K).
   Using IOTA, we also obtained first interferometric observations
   on {\torid}, finding some evidence for a resolved structure, maybe by a
   faint, close companion. }
 {We find indications that the companion C2 is massive itself,
   which makes it likely that its contribution to the intense UV radiation
   field of the Trapezium cluster is non-negligible.
   Furthermore, the high eccentricity of the preliminary orbit solution
   predicts a very small physical separation during periastron passage
   ($\sim1.5$~AU, next passage around 2007.5), suggesting strong wind-wind
   interaction between the two O stars.
 }


   \maketitle
%

\section{Introduction}

Stellar mass is the most fundamental parameter, determining, together with
the chemical composition and the angular momentum, the entire evolution of a 
given star.
Stellar evolutionary models connect these
fundamental parameters with more easily accessible, but also highly uncertain
observables such as the luminosity and the stellar temperature.
Particularly towards the pre-main-sequence (PMS) phase and towards the
extreme stellar masses (i.e.\ the low- and high-mass domain), the existing
stellar evolutionary models are still highly uncertain and require further
empirical verification through direct and unbiased mass estimates, such as
those provided by the dynamical masses accessible in binary systems. 
Recently, several studies were able to provide dynamical masses for low-mass
PMS stars \citep[e.g.\ ][]{tam02, sch03b, bod05}, while direct mass measurements
for young O-type stars are still lacking. 

Furthermore, in contrast to the birth of low-mass stars, the formation
mechanism of high-mass stars is still poorly understood.  In particular, the
remarkably high binary frequency measured for young high-mass stars might
indicate that the way high-mass stars are born differs significantly from the
mass accretion scenario via circumstellar disks, which is well-established for
low- and intermediate-mass stars.  For instance, studies conducted at the 
nearest high-mass star-forming region, the Orion Nebular Cluster \citep[ONC,
at a distance of $440 \pm 34$~pc, ][]{jef07}, revealed 1.1 companions per
primary \citep[for high-mass stars $M > 10 M_{\sun}$,~][]{pre99}, which is
significantly higher than the mean number of companions for intermediate
and low-mass stars.

In the very center of the ONC, four OB stars form the Orion Trapezium; three
of which ($\theta^1$Ori\,A,~B,~C) are known to be multiple \citep{wei99,sch03a}. 
{\torid} (alias HD\,37023, HR\,1896, Parenago~1889) has no confirmed companion,
although a preliminary analysis of the radial velocity by \citet{vit02b}
suggests that it might be a spectroscopic binary with a period of $\sim20$ or
$40$~days.

A particularly intruiging young \citep[$<1$~Myr, ][]{hil97} high-mass star in
the Trapezium cluster is {\toric} (alias 41\,Ori\,C, HD\,37022, HR\,1895,
Parenago~1891). {\toric} is the brightest source within the ONC and also the
main source of the UV radiation ionizing the proplyds and the \object{M42}
\HII~region. A close ($33$~mas) companion with a near-infrared flux ratio of
$\sim0.3$ between the primary ({\toric}1) and the secondary ({\toric}2) 
was discovered in 1997 using bispectrum speckle interferometry
\citep{wei99}.  \citet{don02} estimated the mass of
{\toric} to be $44 \pm 5 M_{\sun}$, making it the most massive star in the
cluster. The same authors give an effective temperature of $45\,000 \pm
1\,000$~K and a stellar radius of $8.2 \pm 1.1~R_{\sun}$. 
\citet{sim06} estimated the mass independently using evolutionary tracks
and by performing a quantitative analysis of {\toric} spectra and obtained
$M_{\rm evol}=33 M_{\sun}$ and $M_{\rm spec}=45 \pm 16 M_{\sun}$,
respectively.
Long series of optical and UV spectroscopic observations revealed that the
intensity and also some line profiles vary in a strictly periodic way.  With
$15.422 \pm 0.002$~days, the shortest period was reported by \citet{sta93}.
Several authors interpret this periodicity, which in the meantime was also
detected in X-ray \citep{gag97}, within an oblique magnetic rotator model,
identifying 15.422~d with the rotation period of the star. \citet{sta96}
detected a steady increase in radial velocity, confirmed by \citeauthor{don02}
in 2002, which suggests a spectroscopic binary with an orbital period of at
least 8~years. \citet{vit02} searched for long-term periodicity in the radial 
velocity and reported two additional periods of 66~days and 120~years, which
he interpreted as the presence of, in total, three components in the system.  

Given the unknown orbit of the speckle companion, it still must be
determined which one of these periods corresponds to the orbital motion of C2.
Since the discovery of C2 in 1997, three measurements performed with bispectrum speckle
interferometry showed that the companion indeed undertakes orbital motion
\citep{sch03a}, reaching the largest separation of the two components in autumn
1999 with $43 \pm 2$~mas.  In order to follow the orbital motion, we
monitored the system using infrared and visual bispectrum speckle
interferometry and in 2005, for the first time, also using infrared
long-baseline interferometry.

An interesting aspect of the dynamical history of the ONC was presented
by~\citet{tan04}.  He proposed that the Becklin-Neugebauer (BN) object, which
is located 45\arcsec\ northwest of the Trapezium stars, might be a
runaway B star ejected from the {\toric} multiple system approximately
$4\,000$~yrs ago.  This scenario is based on proper motion measurements, which
show that BN and {\toric} recoil roughly in opposite directions, and by the
detection of X-ray emission potentially tracing a wind bow
shock\footnote{However, the more recent detection of X-ray variability in
  intensity and spectrum makes it unlikely that this X-ray emission
  really originates in a wind bow shock, as pointed out by \citet{gro05}.}.
Three-body interaction is a crucial part of this interpretation, and C2 is
currently the only candidate which could have been involved. Therefore, a
high-precision orbit measurement of C2 might offer the unique possibility to
recover the dynamical details of this recent stellar ejection.
However, another study \citep{rod05} also aimed to identify the multiple
system from which BN was ejected and identified Source I as the likely
progenitor system.  Later, \citet{gom05} added further evidence to this
interpretation by identifying Source $n$ as a potential third
member of the decayed system.

\section{Observations and Data Reduction}

\begin{table*}[t]
\caption{Observation Log.}
\label{tab:observations}
\centering

\begin{tabular}{llcclcl}
  \hline\hline
  Target & Instrument           & Date        & Filter$^{a}$ & Detector$^{b}$ & No. Interferograms & Calibrators$^{d}$\\
         & and Configuration    & [UT]        &              & and Mode$^{c}$ & Target/Calibrator & \\
  \noalign{\smallskip}
  \hline
  \noalign{\smallskip}
  \toric & BTA~6m/Speckle     & 1997.784    & $H$      & P/DIT=150\,ms & 519/641            & {\torid}\\
  \toric & BTA~6m/Speckle     & 1998.838    & $K'$     & H/DIT=120\,ms & 438/265            & {\torid}\\
  \toric & BTA~6m/Speckle     & 1999.737    & $J$      & H/DIT=100\,ms & 516/244            & {\torid}\\
  \toric & BTA~6m/Speckle     & 1999.8189   & $G'$     & S/DIT=5\,ms   & 500/--             & --\\
  \toric & BTA~6m/Speckle     & 2000.8734   & $V'$     & S/DIT=5\,ms   & 1\,000/--          & --\\
  \toric & BTA~6m/Speckle     & 2001.184    & $J$      & H/DIT=80\,ms  & 684/1\,523         & {\torid}\\
  \toric & BTA~6m/Speckle     & 2003.8      & $J$      & H/DIT=160\,ms & 312/424            & {\torid}\\
  \toric & BTA~6m/Speckle     & 2003.9254   & $V'$     & S/DIT=2.5\,ms & 1\,500/--          & --\\
  \toric & BTA~6m/Speckle     & 2003.928    & $V'$     & S/DIT=2.5\,ms & 2\,000/--          & --\\
  \toric & BTA~6m/Speckle     & 2004.8216   & $V'$     & S/DIT=5\,ms   & 2\,000/--          & --\\
  \toric & BTA~6m/Speckle     & 2006.8      & $V'$, $R'$ & --            & --                 & --\\
  \hline
  \toric & IOTA A35-B15-C0    & 2005 Dec 04 & $H$      & 1L7R, 2L7R  & 11\,400/8\,050       & HD\,14129, HD\,36134, HD\,34137,\\
                   &                    &              &        &             &                    & HD\,50281, HD\,63838\\
  \toric & IOTA A35-B15-C10   & 2005 Dec 02 & $H$      & 2L7R, 4L7R  & 4\,400/4\,950        & HD\,34137, HD\,50281, HD\,63838\\
  \toric & IOTA A35-B15-C10   & 2005 Dec 03 & $H$      & 2L7R        & 4\,600/2\,450        & HD\,28322\\
  \toric & IOTA A35-B15-C15   & 2005 Dec 01 & $H$      & 2L7R, 4L7R  & 7\,250/5\,000        & HD\,20791, HD\,34137, HD\,36134\\
  \toric & IOTA A25-B15-C0    & 2005 Dec 06 & $H$      & 2L7R, 4L7R  & 5\,250/4\,875        & HD\,28322, HD\,34137, HD\,36134,\\
                   &                    &              &        &             &                    & HD\,74794\\
  \hline
  \torid & IOTA A35-B15-C0    & 2005 Dec 04 & $H$      & 2L7R        & 800/2\,800           & HD\,14129, HD\,36134, HD\,34137,\\
                   &                    &              &        &             &                    & HD\,50281, HD\,63838\\
  \torid & IOTA A25-B15-C0    & 2005 Dec 06 & $H$      & 2L7R, 4L7R  & 1\,800/4\,875        & HD\,28322, HD\,34137, HD\,36134,\\
                   &                    &              &        &             &                    & HD\,74794\\
  \noalign{\smallskip}
  \hline
\end{tabular}

\begin{flushleft}
  \hspace{5mm}Notes~--~$^{a)}$ Filter central wavelength and bandwidth, in nm ($\lambda_{c}$/$\Delta\lambda$)~--~$V'$:~545/30; $G'$:~610/20; $R'$:~800/60; $J$:~1\,239/138; $H$:~1\,613/304; $K'$:~2\,115/214.\\
  \hspace{5mm}$^{b)}$~P: PICNIC detector, H: HAWAII array, S: Multialkali
  S25 intensifier photocathode\\
  \hspace{5mm}$^{c)}$~For the IOTA measurements, we used different
  detector read modes to adapt to the changing atmospheric conditions.
  The two numbers in the given 4-digit code denote the value of the {\it
    loop} and {\it read} parameter \citep{ped04} of the PICNIC camera.
  Since data taken in different readout modes is calibrated independently, the
  scattering between the data sets also resembles the typical calibration errors
  (see Figure~\ref{fig:fitC}).\\
  \hspace{5mm}$^{d)}$~The dash symbol in the calibrator column
    indicates speckle measurements for which no calibrator was observed.\\
\end{flushleft}
\end{table*}

\begin{table}[t]
\caption{IOTA Calibrator Stars Information}
\label{tab:calibrators}
\centering

\begin{tabular}{lcccc}
  \hline\hline
  Star                & V & H & Spectral & Adopted UD diameter\\
                      &   &   & Type     & [mas]\\
  \noalign{\smallskip}
  \hline
  \noalign{\smallskip}
  \object{HD\,14129}  & 5.5 & 3.3 & G8.5III & $1.01 \pm 0.01$$^{a}$\\
  \object{HD\,20791}  & 5.7 & 3.5 & G8.5III & $0.89 \pm 0.01$$^{a}$\\
  \object{HD\,28322}  & 6.2 & 3.9 & G9III   & $0.82 \pm 0.01$$^{a}$\\
  \object{HD\,34137}  & 7.2 & 4.4 & K2III   & $0.80 \pm 0.01$$^{a}$\\
  \object{HD\,36134}  & 5.8 & 3.2 & K1III   & $1.16 \pm 0.02$$^{a}$\\
  \object{HD\,50281}  & 6.6 & 4.3 & K3V     & $0.77 \pm 0.10$$^{b}$\\
  \object{HD\,63838}  & 6.4 & 3.6 & K2III   & $0.95 \pm 0.01$$^{a}$\\
  \object{HD\,74794}  & 5.7 & 3.5 & K0III   & $1.07 \pm 0.01$$^{a}$\\
  \noalign{\smallskip}
  \hline
\end{tabular}

\begin{flushleft}
  \hspace{5mm}Notes~--~$^{a}$ UD diameter taken from the CHARM2 catalog~\citep{ric05}.\\
  \hspace{5mm}$^{b}$ UD diameter taken from getCal tool {\tt (http://mscweb.ipac.caltech.edu/gcWeb/gcWeb.jsp)}.\\
\end{flushleft}
\end{table}

\subsection{Bispectrum speckle interferometry}

Speckle interferometric methods are powerful techniques for overcoming the atmospheric
perturbations and for reaching the diffraction-limited resolution of ground-based
telescopes, both at near-infrared and visual wavelengths.
Since the discovery of {\toric}2 in 1997, we have monitored the system
with the \textit{Big Telescope Alt-azimuthal} (BTA) 6.0~m telescope of the
Special Astrophysical Observatory located on Mt. Pastukhov in Russia. 
For the speckle observations at visual wavelengths, a $1280\times1024$ pixel
CCD with a multialkali S25 intensifier photocathode was used.
The near-infrared speckle observations were carried out using one
512$\times$512 pixel quadrant of the Rockwell HAWAII array in our speckle
camera, with pixel sizes of 13.4~mas ($J$-band), 20.2~mas ($H$-band), and
27~mas ($K$-band) on the sky.

For the speckle observations at infrared wavelengths, we recorded
interferograms of {\toric} and of the nearby unresolved star {\torid}~in order
to compensate for the atmospheric speckle transfer function.  The number of
interferograms and the detector integration times (DITs) are listed in
Table~\ref{tab:observations}. 

The modulus of the Fourier transform of the object (visibility) was obtained
with the speckle interferometry method~\citep{lab70}.  For image
reconstruction we used the bispectrum speckle interferometry method
(\citealt{wei77}, \citealt{wei83}, \citealt{loh83}, \citealt{hof86}).

\subsection{IOTA long-baseline interferometry}

\begin{figure}[tbp]
  \centering
  \includegraphics[width=8.5cm]{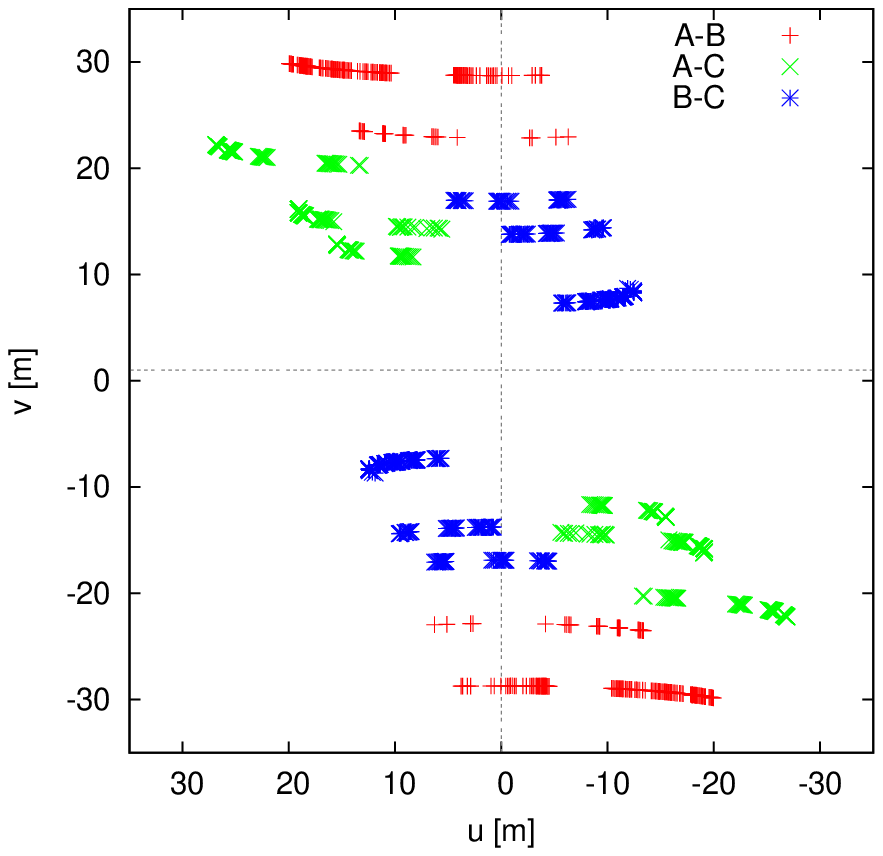}\\
  \caption{$uv$-plane coverage obtained on
  \textbf{\boldmath\toric\boldmath}~with the four IOTA array
  configurations used.  The relatively strong asymmetry in the $uv$-plane
  coverage mainly results from the position of {\toric} relatively
  close to the celestrial equator. }
  \label{fig:iotauvcovC}
\end{figure}

\begin{figure}[tbp]
  \centering
  \includegraphics[width=8.5cm]{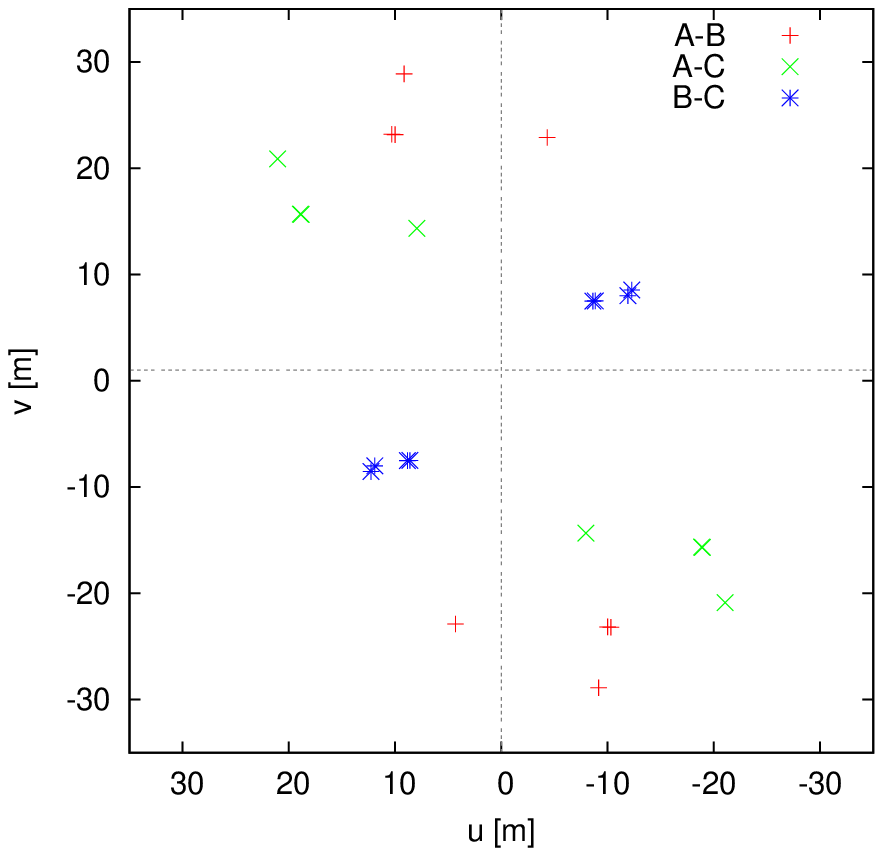}\\
  \caption{$uv$-plane coverage obtained on
    \textbf{\boldmath\torid\boldmath}~with two IOTA array configurations.
    Symbols and colors as in Figure~\ref{fig:iotauvcovC}.
  }
  \label{fig:iotauvcovD}
\end{figure}

The \textit{Infrared Optical Telescope Array} (IOTA) is a three-telescope,
long-baseline interferometer located at the Fred Lawrence Whipple Observatory on
Mount Hopkins, Arizona, operating at visual and near-infrared
wavelengths~\citep{tra03}. Its three 45~cm primary Cassegrain telescopes can
be mounted on stations along an L-shaped track, reaching 15~m towards a
southeastern and 35~m towards a northeastern direction.
After passing a tip-tilt system, which compensates the atmospherically
induced motion of the image, and path-compensating delay lines, the three
beams are fed into fibers and coupled pairwise onto the IONIC3 integrated
optics beam combiner~\citep{ber03}.
The interferograms are recorded by temporal modulation around zero optical
path delay (OPD).  During data acquisition, a fringe tracker
software~\citep{ped05} continuously compensates potential OPD drifts.  This
allows us to measure the three interferograms nearly simultaneously within the
atmospheric coherence time, preserving the valuable closure phase (CP)
information. 

For our IOTA observations, we used four different array configurations (see
Table~\ref{tab:observations}), obtaining the $uv$-coverage shown in
Figure~\ref{fig:iotauvcovC}. {\torid} was observed on two different array
configurations, as shown in Figure~\ref{fig:iotauvcovD}.
During each night, we systematically alternated between the target star and
calibrators in order to determine the transfer function of the instrument.
For more details about the calibrator stars and the number of recorded
Michelson interferograms, refer to Table~\ref{tab:calibrators}.

In order to extract visibilities and CPs from the recorded IOTA
interferograms, we used the IDRS\footnote{The IDRS data reduction software can
  be obtained from {\tt ftp://ftp.mpifr-bonn.mpg.de/outgoing/skraus/idrs/}} data reduction
software.  Basic principles of the algorithms implemented in this software
package were already presented in \citet{kra05}, although several details have
been refined to obtain optimal results for fainter sources as well, such as
those observed in this study. 

To estimate the fringe amplitude (visibility squared, $V^2$), we
compute the continuous wavelet transform (CWT), which decomposes the signal
into the OPD-scale domain, providing scale (frequency) resolution while
preserving the information about the fringe OPD.
The turbulent Earth atmosphere introduces fast-changing OPD variations
between the combined telescopes (also known as atmospheric piston),
degenerating the recorded interferograms.  By measuring the extension of the
fringe packet in the CWT along both the scale and the OPD axis, we identify
the scans which are most affected by this effect and reject them from
further processing.
For the remaining scans, we apply a method similar to the procedure presented
by \citet{ker04}.  First, the fringe peak is localized in the CWT.  In order
to minimize noise contributions, a small window around the fringe peak
position is cut out.  Then we integrate along the OPD axis, yielding a power
spectrum.  After recentering the fringe peak position for each scan (to
compensate frequency changes induced by atmospheric piston), we average the
power spectra for all scans within a dataset.  In the resulting averaged power
spectrum, we fit and remove the background contributions and integrate over
the fringe power to obtain an estimate for $V^2$.

Another refinement in our software concerns the CP estimation.
We found that the best signal-to-noise ratio (SNR) can be achieved by
averaging the bispectra from all scans.  The bispectrum is given by the triple
product of the Fourier transform of the scans at the three baselines
\citep{hof93}.  Then, we use the triple amplitude to select the bispectrum
elements with the highest SNR and average the triple phases of these elements
in the complex plane to obtain the average closure phase.

\section{Aperture synthesis imaging}\label{sec:imgrec}

\begin{figure*}[tbp]
  \centering
  \begin{tabular}{c}
    \begin{minipage}{20cm}
      $\begin{array}{ccc}
        \hspace{-5mm}\textnormal{{\bf 1999.7}, BTA, $J$-band} &
        \hspace{-18mm}\textnormal{{\bf 2004.8}, BTA, $V'$-band} &
        \hspace{-18mm}\textnormal{{\bf 2005.9}, IOTA, $H$-band}\\[-11mm]
        \hspace{-8mm}\includegraphics[width=7.3cm]{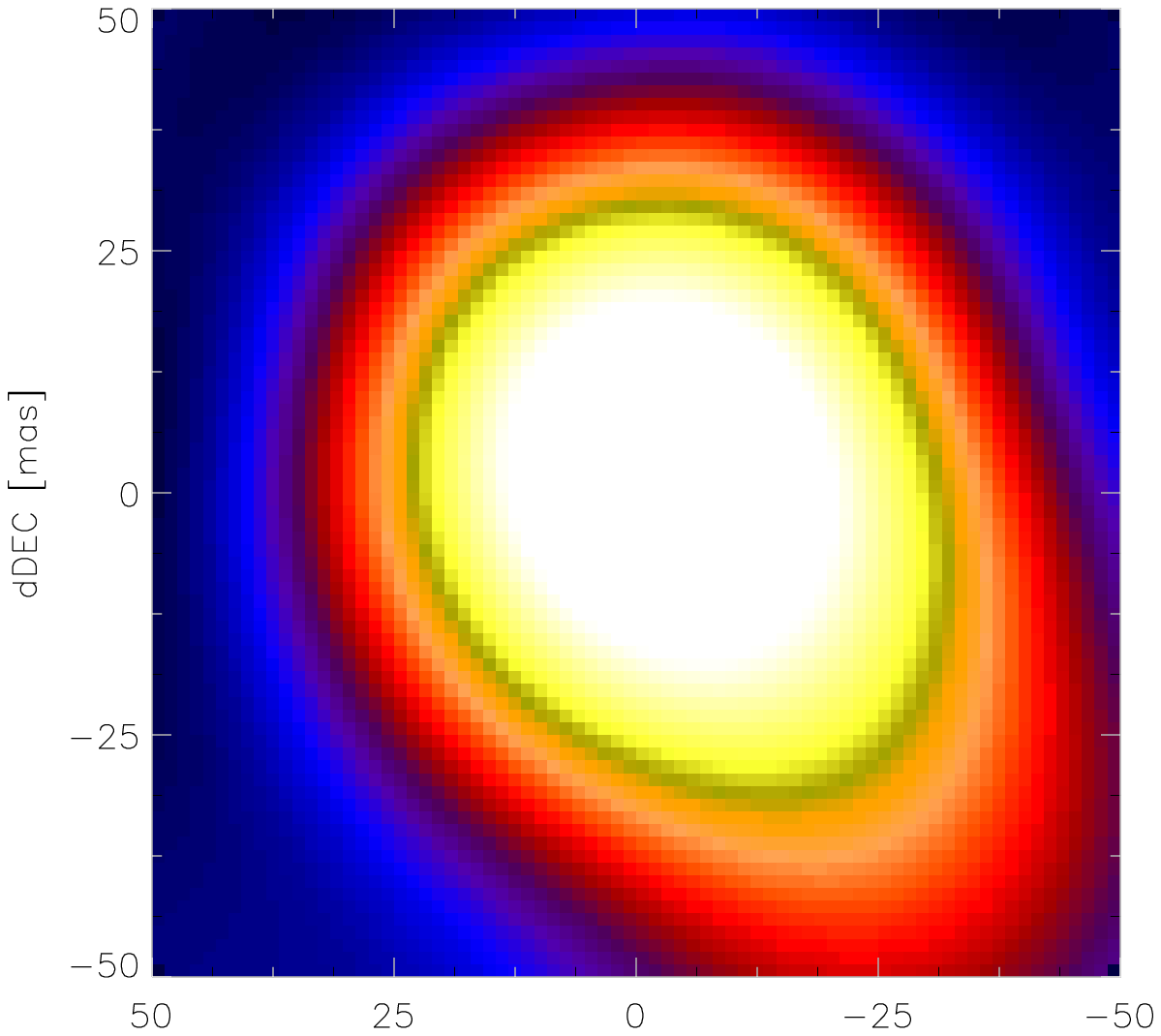} & \hspace{-2cm}\includegraphics[width=7.3cm]{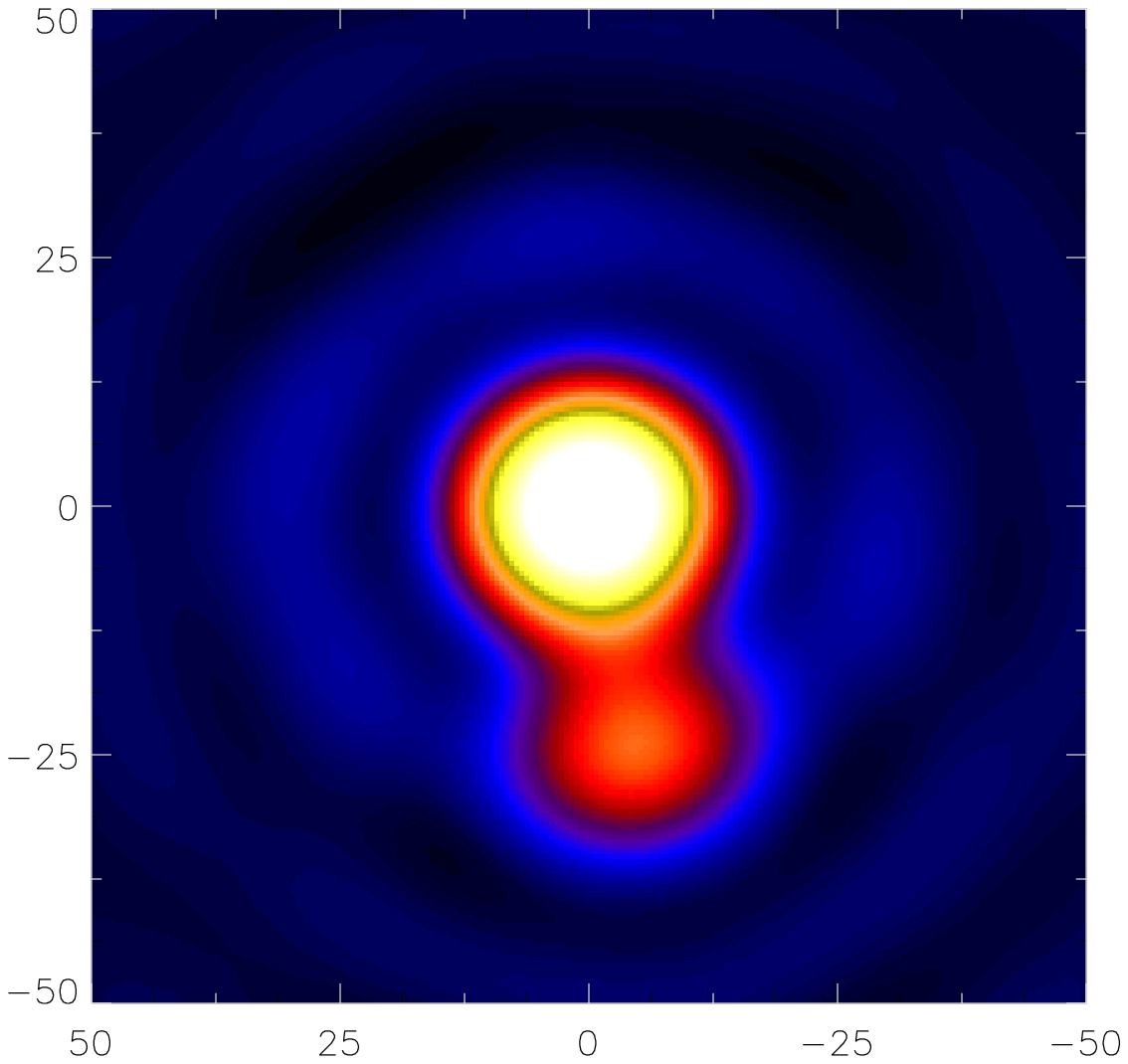} & \hspace{-2cm}\includegraphics[width=7.3cm]{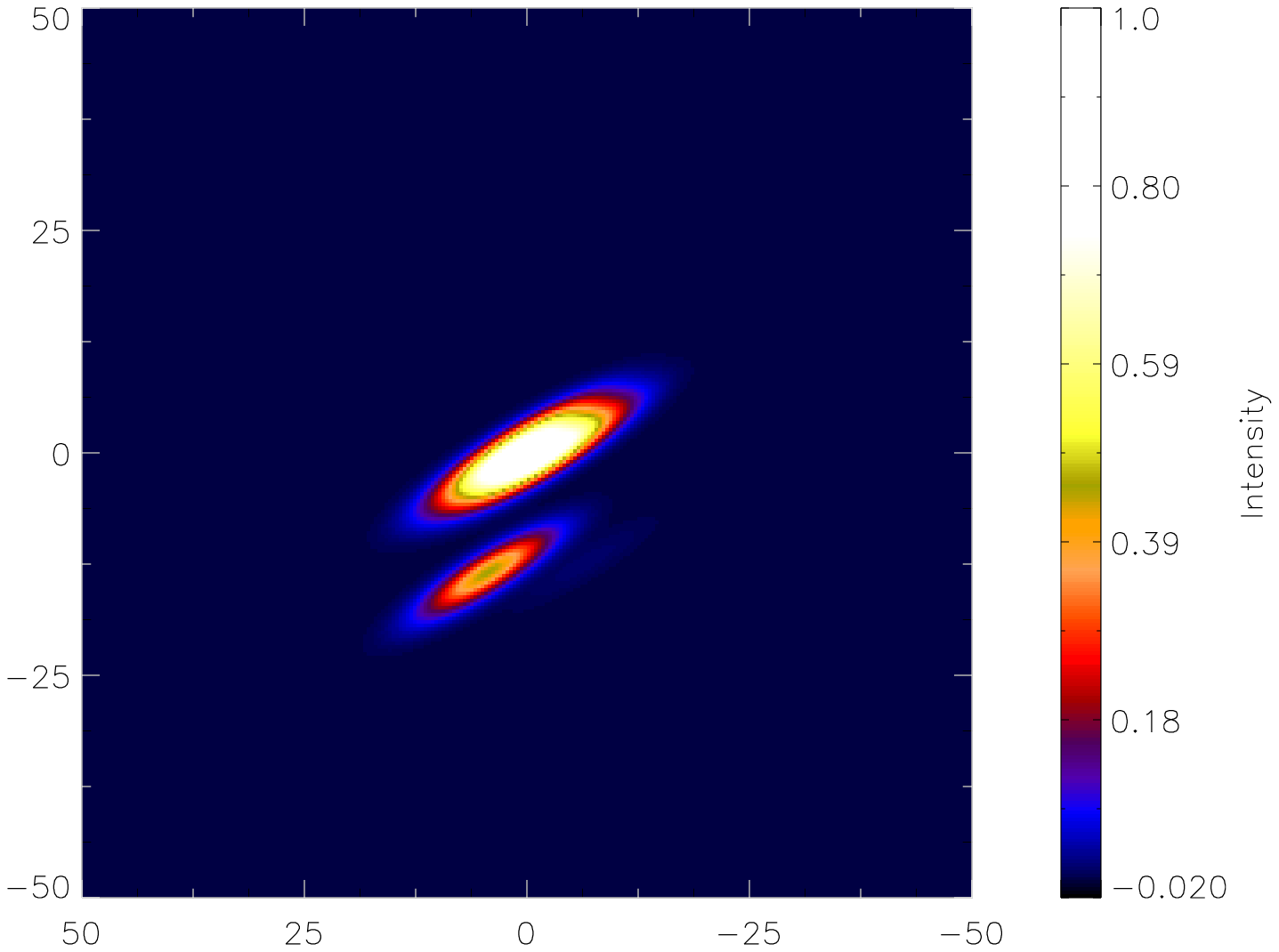}\\[-21mm]
        \hspace{-8mm}\includegraphics[width=7.3cm]{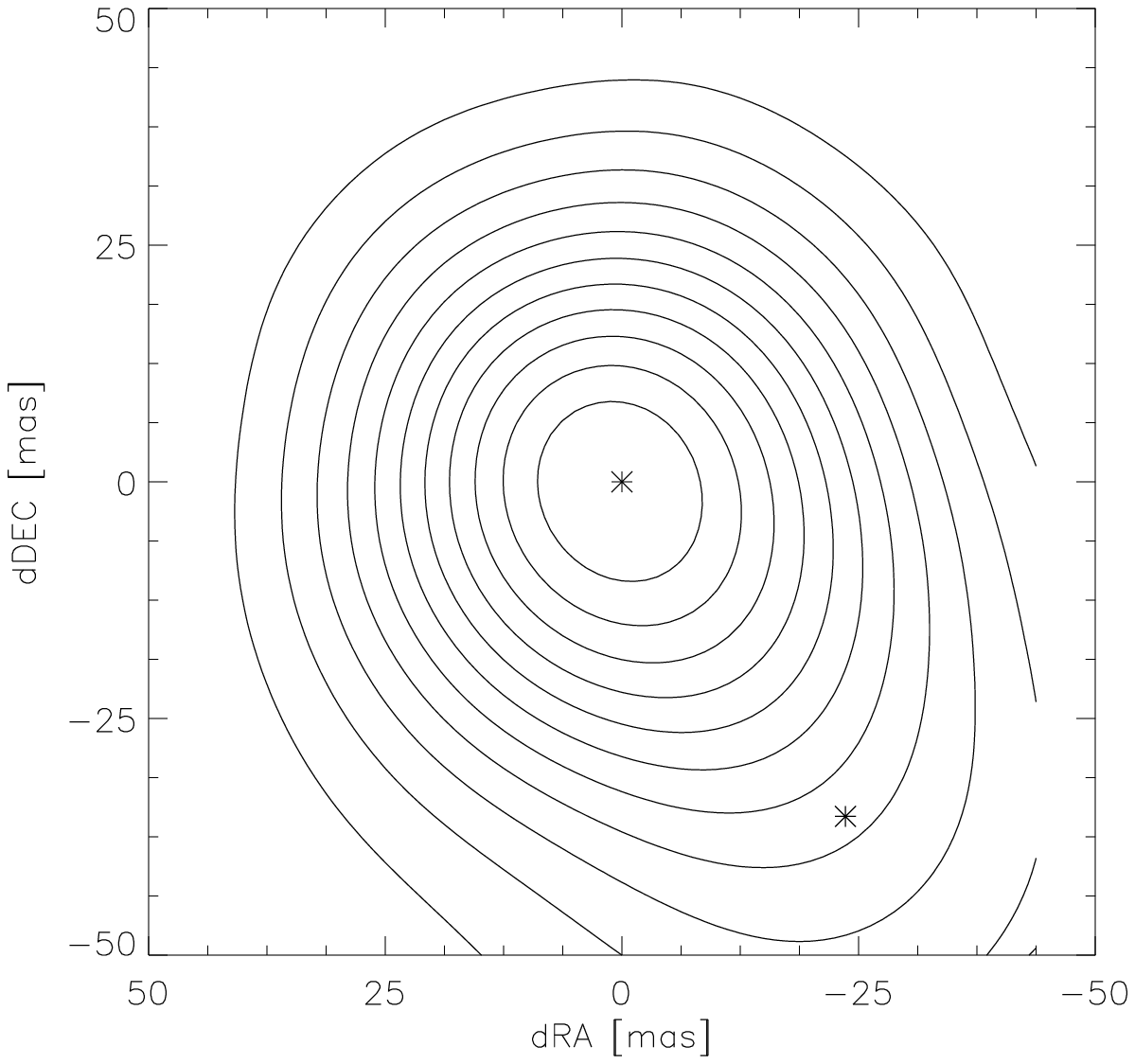} & \hspace{-2cm}\includegraphics[width=7.3cm]{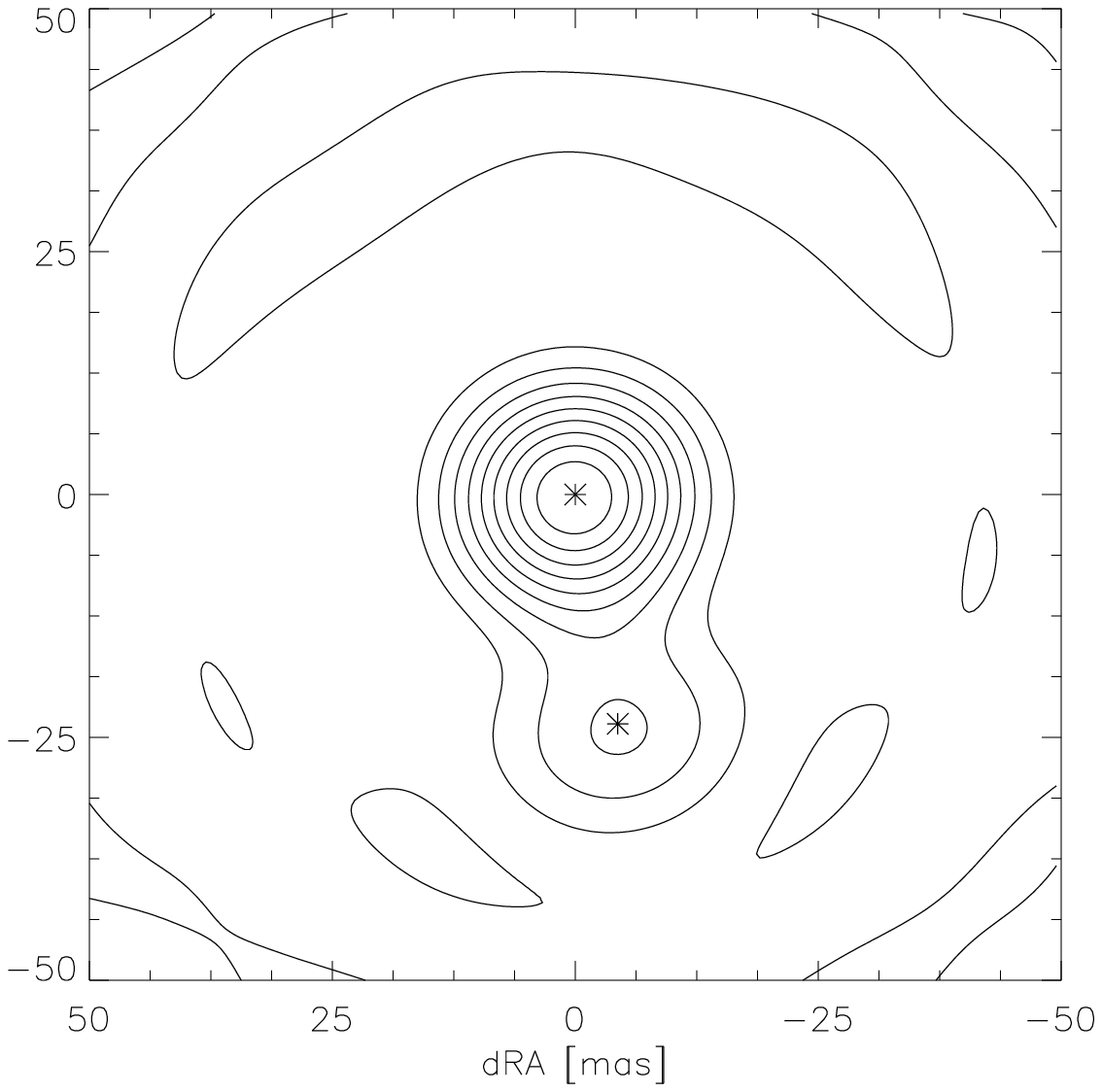} & \hspace{-2cm}\includegraphics[width=7.3cm]{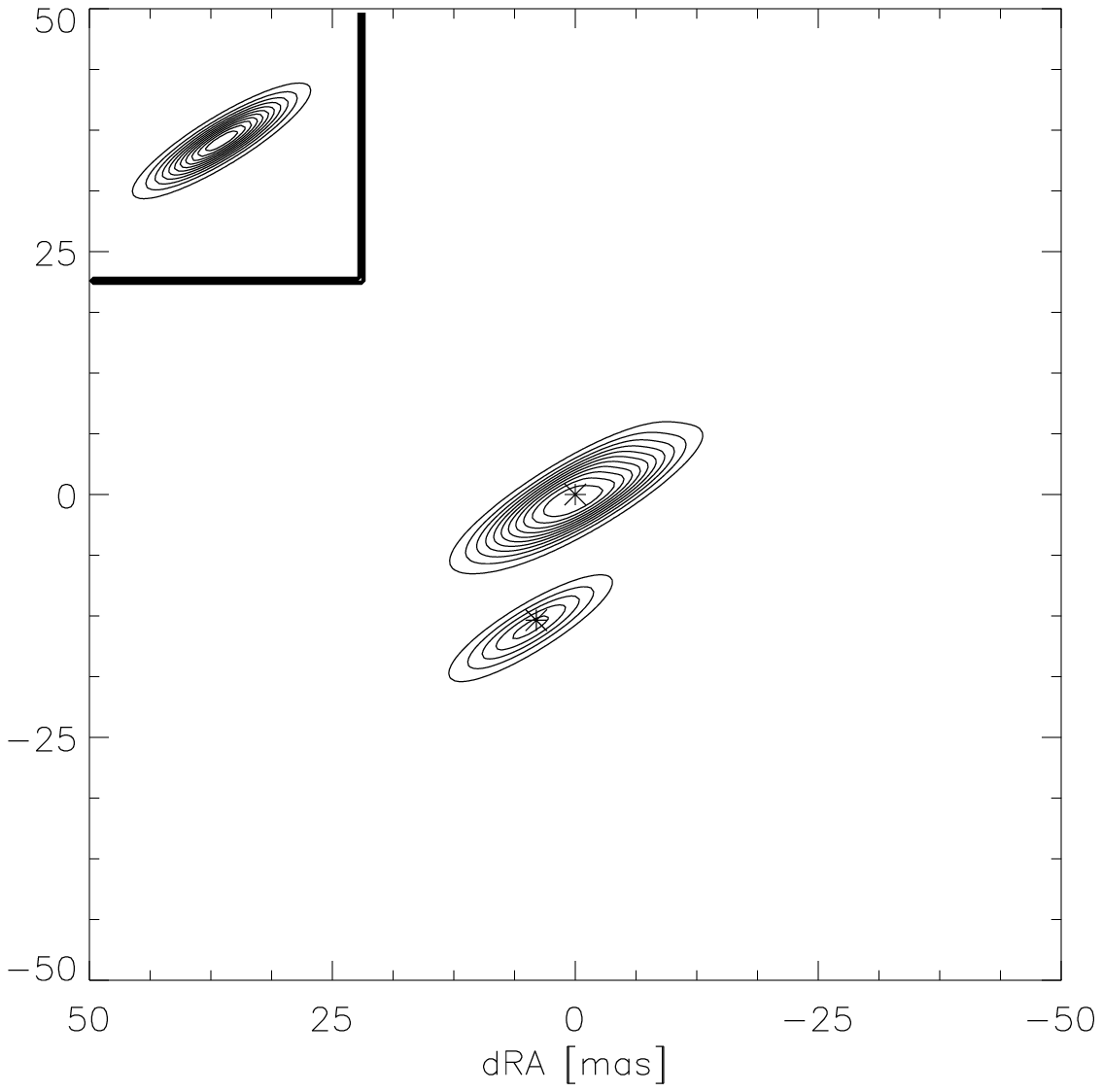}
      \end{array}$
    \vspace{-1cm}
    \end{minipage}
  \end{tabular}
  \caption{\textit{Left, Middle:} Bispectrum speckle $J$ and $V'$-band images
    reconstructed from BTA-data. \textit{Right:} Aperture synthesis image of
    {\toric} reconstructed from our IOTA $H$-band data.
    Besides the false-color representation in the upper row, we show the images
    below as contours with the best-fitted positions marked with star symbols
    (see Table~\ref{tab:position}).  In the image in the lower right corner,
    the restoring beam for the IOTA aperture synthesis image is shown  as an
    inset.
    Over the six year interval covered by the 
    images, orbital motion is clearly conceivable (1999/2004/2005:
    $\rho$=43/24/14~mas; $\Theta$=214\degr/191\degr/163\degr). }
  \label{fig:image}
\end{figure*}

Interpreting optical long-baseline interferometric data often
requires a priori knowledge about the expected source brightness
distribution.  This knowledge is used to choose an astrophysically motivated
model whose parameters are fitted to the measured interferometric
observables (as applied in Sect.~\ref{sec:iotamodelfit}).

However, the measurement of CPs allows a much more intuitive
approach; namely, the direct reconstruction of an aperture synthesis image.
Due to the rather small number of telescopes combined in the current
generation of optical interferometric arrays, direct image
reconstruction is limited to objects with a rather simple source geometry; in
particular, multiple systems \citep[for images reconstructed from IOTA data,
see][]{mon04, kra05}. 

Using our software based on the \textit{Building Block Mapping}
algorithm~\citep{hof93}, we reconstructed an aperture synthesis image of
the {\toric} system from the data collected during our IOTA run.
Starting from an initial single delta-function, this algorithm builds up the
image iteratively by adding components in order to minimize the least-square
distance between the measured bispectrum and the bispectrum of the
reconstructed image.

The resulting image is shown in Figure~\ref{fig:image} and provides a
model-independent representation of our data. 
By combining the data collected during six nights, we make the
reasonable assumption that the orbital motion over this interval is
negligible.

The clean beam, which we used for convolution to obtain the final image, is
rather elliptical (see inset in Figure~\ref{fig:image}), representing the
asymmetries in the $uv$-coverage.

\section{Model fitting} \label{sec:iotamodelfit}

\subsection{Binary model fitting for \toric} \label{sec:t1oricmodel}

\begin{figure*}[tbp]
  \centering
  \includegraphics[width=17cm]{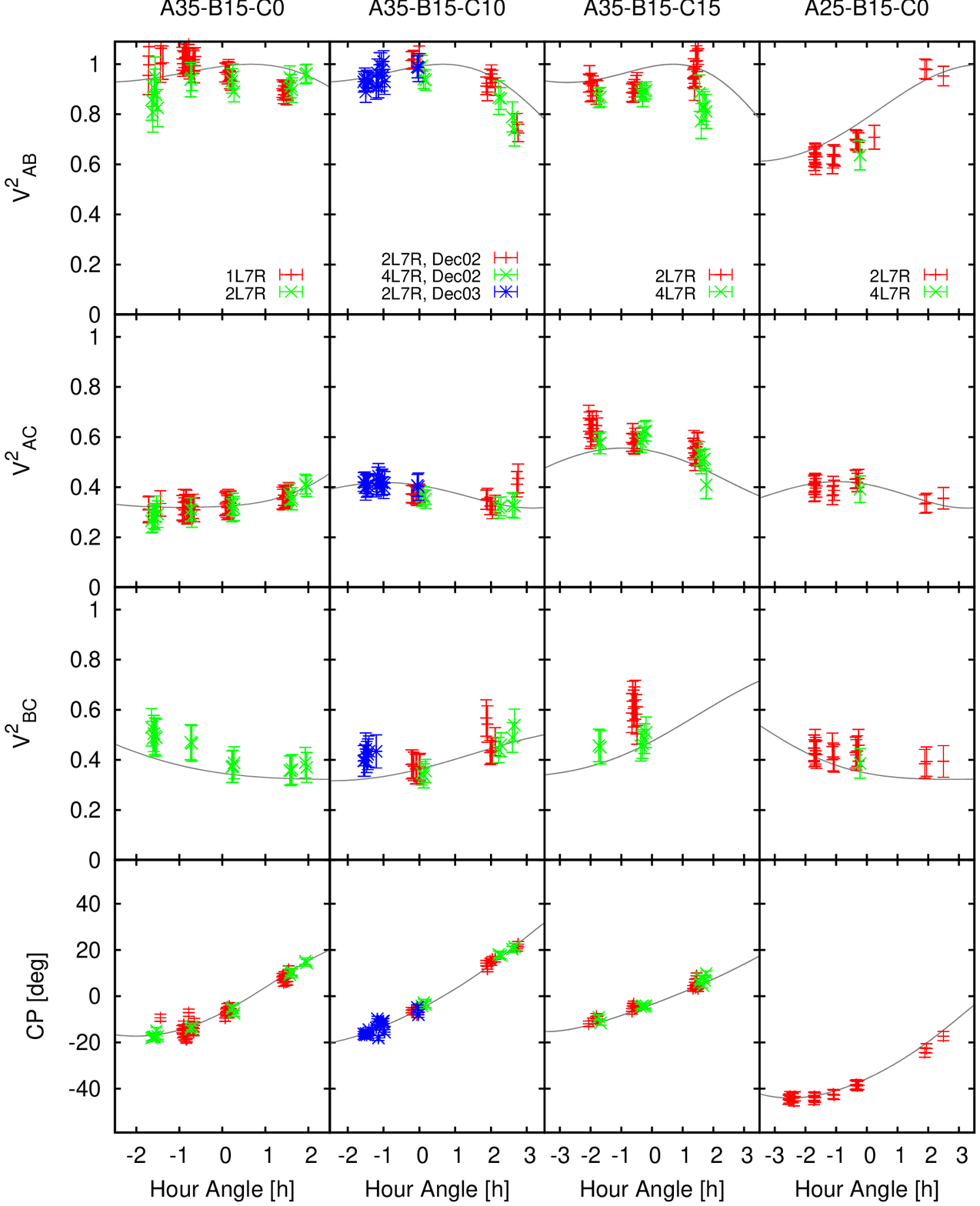}\\
  \caption{Visibilities and Closure Phases derived from the recorded IOTA
  interferograms for \textbf{\boldmath\toric\boldmath}~as a function of hour angle (points with error bars). 
  The solid line shows the binary model fit, described in more detail 
  in Section~\ref{sec:iotamodelfit}.  The different symbols represent the
  different detector modes used (see Table~\ref{tab:observations}). The data
  for each detector mode was calibrated separately, so the scattering of the
  data groups represents the typical calibration errors.
}
  \label{fig:fitC}
\end{figure*}

\begin{figure}[tbp]
  \centering
  \includegraphics[width=8.5cm]{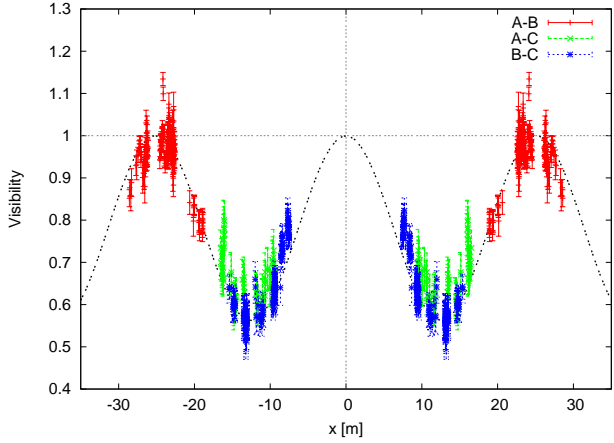}\\
  \caption{Projection of the sampled Fourier plane along the fitted binary PA
  of $162.74\degr$ ($x=u\cos\Theta - v\sin\Theta$), clearly revealing the
  binary signature.  The dashed line shows the theoretical cosine visibility
  profile for a binary star with separation 13.55~mas and intensity ratio
  0.28.}
  \label{fig:projectuv}
\end{figure}

\begin{table*}[t]
\caption{Relative astrometry and photometry for the {\toric} binary system}
\label{tab:position}
\centering
\begin{tabular}{lllccccccccc}
  \hline\hline
                         &                &        &                 &                 &             &      & \multicolumn{2}{c}{(O--C) Orbit \#1} & & \multicolumn{2}{c}{(O--C) Orbit \#2}\\
                         \cline{8-9} \cline{11-12}
  Telescope              & Date           & Filter & Flux ratio      & $\Theta^a$      & $\rho$      & Ref. & $\Theta$       & $\rho$       & & $\Theta$       & $\rho$      \\
                         &                &        & $F_{\rm C2}/F_{\rm C1}$ & [\degr] & [mas]       &      & [\degr]        & [mas]        & & [\degr]        & [mas]       \\
  \noalign{\smallskip}
  \hline
  \noalign{\smallskip}
  BTA~6m/Speckle   & 1997.784             & H      & $0.26 \pm 0.02$ & $226.0 \pm 3$   & $33 \pm 2$      & b    & +3.0 & +0.0  && +3.0 & +0.5 \\
  BTA~6m/Speckle   & 1998.838             & K'     & $0.32 \pm 0.03$ & $222.0 \pm 5$   & $37 \pm 4$      & b    & +3.8 & -2.6  && +3.8 & -2.5 \\
  BTA~6m/Speckle   & 1999.737             & J      & $0.31 \pm 0.02$ & $214.0 \pm 2$   & $43 \pm 1$      & c    & -0.9 & +1.5  && -0.9 & +1.5 \\
  BTA~6m/Speckle   & 1999.8189            & G'     & $0.35 \pm 0.04$ & $213.5 \pm 2$   & $42 \pm 1$      & --   & -1.1 & +0.5  && -1.1 & +0.5 \\
  BTA~6m/Speckle   & 2000.8734            & V'     & $0.35 \pm 0.03$ & $210.0 \pm 2$   & $40 \pm 1$      & --   & -0.8 & -0.8  && -0.9 & -0.8 \\
  BTA~6m/Speckle   & 2001.184             & J      & $0.29 \pm 0.02$ & $208.0 \pm 2$   & $38 \pm 1$      & c    & -1.6 & -2.1  && -1.7 & -2.1 \\
  BTA~6m/Speckle   & 2003.8               & J      & $0.30 \pm 0.02$ & $199.3 \pm 2$   & $29 \pm 2$      & --   & +3.9 & +0.5  && +2.8 & +0.5 \\
  BTA~6m/Speckle   & 2003.9254            & V'     & --              & $199.0 \pm 2$   & $29 \pm 2$      & --   & +3.6 & +1.3  && +3.4 & +1.3 \\
  BTA~6m/Speckle   & 2003.928             & V'     & --              & $199.1 \pm 2$   & $29 \pm 2$      & --   & +3.8 & +1.3  && +3.6 & +1.3 \\
  BTA~6m/Speckle   & 2004.8216            & V'     & $0.34 \pm 0.04$ & $190.5 \pm 4$   & $24 \pm 4$      & --   & +4.2 & +2.4  && +4.0 & +2.4 \\
  IOTA             & 2005.92055           & H      & $0.28 \pm 0.03$ & $162.74 \pm 2$  & $13.55 \pm 0.5$ & --   & -0.5 & +0.0  && -1.0 & +0.0 \\
  BTA~6m/Speckle   & 2006.8               & V', R  & --              & --              & $<15$           & --   & --   & --    && --   & --   \\
  \noalign{\smallskip}
  \hline
\end{tabular}
\begin{flushleft}
  \hspace{11mm}Notes~--~$^{a)}$~Following the convention, we measure the position angle (PA) from north to east.\\
  \hspace{5mm}References~--~$^{b)}$~\citealt{wei99}, $^{c)}$~\citealt{sch03a}\\
\end{flushleft}
\end{table*}

Although the aperture synthesis image presented in the last section might also
be used to extract parameters like binary separation, orientation, and
intensity ratio of the components ($I_{\rm C2}/I_{\rm C1}=0.26$, $\rho=14.1$~mas,
$\Theta=162$\degr), more precise values, including error estimates, can be obtained by
fitting the measured visibilities and CPs to an analytical binary model.

The applied model is based on equations 7--12 presented in \citet{kra05} and
uses the least-square Levenberg-Marquardt method to determine the best-fit
binary separation vector and intensity ratio.  In order to avoid potential
local minima, we vary the initial values for the least-square fit on a grid,
searching for the global minimum.

Since the apparent stellar diameter of {\toric} is expected to be 
only $\sim0.2$~mas at the distance of Orion, for our fits we assume that both
stellar components appear practically unresolved to the IOTA baselines.
Furthermore, we assume that the relative  position of the components did not
change significantly over the 6~nights of observation.

Figure~\ref{fig:fitC} shows the measured IOTA visibilities and CPs and
the observables corresponding to our best-fit binary model 
($\chi^2_{V^2}/\textnormal{dof}=1.35$, $\chi^2_{CP}/\textnormal{dof}=1.48$).
The separation $\rho$, PA $\Theta$, and intensity ratio of this binary model
are given in Table~\ref{tab:position}, together with the positions derived from
the speckle observations.
To illustrate more clearly that the measured IOTA visibilities resemble a
binary signature, in Figure~\ref{fig:projectuv} we show a projection of the
sampled two-dimensional Fourier plane along the binary vector, revealing the
cosine modulation corresponding to the Fourier transform of a binary
brightness distribution.

For the speckle data (providing a complete Fourier sampling up to the spatial
frequency corresponding to the diameter of the telescope primary mirror), we
determine the binary parameters by fitting a two-dimensional cosine function
directly to the 2-D speckle interferogram power spectrum.

\subsection{Resolved structure around \torid:\\
Potential detection of a companion} \label{sec:t1oridmodel}

\begin{figure}[tbp]
  \centering
  \vspace{2mm}
  \includegraphics[width=8cm]{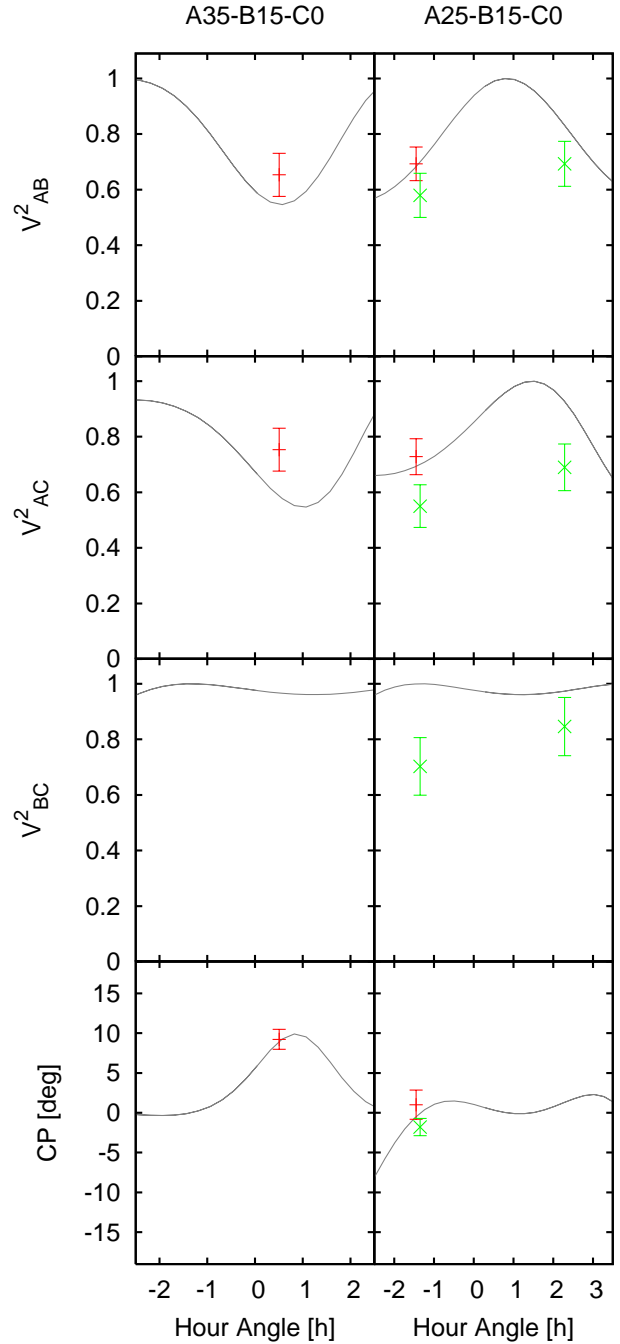}\\[2mm]
  \caption{Visibilities and Closure Phases derived from the recorded IOTA
  interferograms for \textbf{\boldmath\torid\boldmath}~(points with error bars). The solid line shows the
  binary model fit, described in more detail in Section~\ref{sec:t1oridmodel}. }
  \label{fig:fitD}
\end{figure}

Besides the main target of our observational programme, \toric, during the two
 nights with the best seeing conditions, we also recorded four datasets on
\torid.  Despite lower flux (\torid: $H$=5.9, \toric: $H$=4.6), the
quality of the derived visibilities and CPs seems reliable,
although slightly larger errors must be assumed. {\torid} appears resolved in our
measurements with a significant non-zero CPs signal ($\sim10$\degr) on the
A35-B15-C0 baseline.  This CP indicates deviations from point-symmetry, as
expected for a binary star. 
We applied the binary model fit described in Sect.~\ref{sec:t1oricmodel} and
found the binary system with an intensity ratio of 0.14,
 $\rho=18.4$~mas, and $\Theta=41$\degr~(Figure~\ref{fig:fitD}) to be the
 best-fit model ($\chi^2/\textnormal{dof}=1.36$).

However, considering the $uv$-coverage of the existing dataset, this solution
is likely not unique, and it can not be ruled out that other geometries, 
such as for inclined circumstellar disk geometries with pronounced emission
from the rim at the dust sublimation radius \citep[see e.g.\ ][]{mon06},
might also produce the asymmetry required to fit the data.

\section{Results}

\subsection{Preliminary physical orbit of the {\toric} binary system} \label{sec:orbitsolution}

\begin{figure}[tbp]
  \centering
  \includegraphics[width=8.5cm]{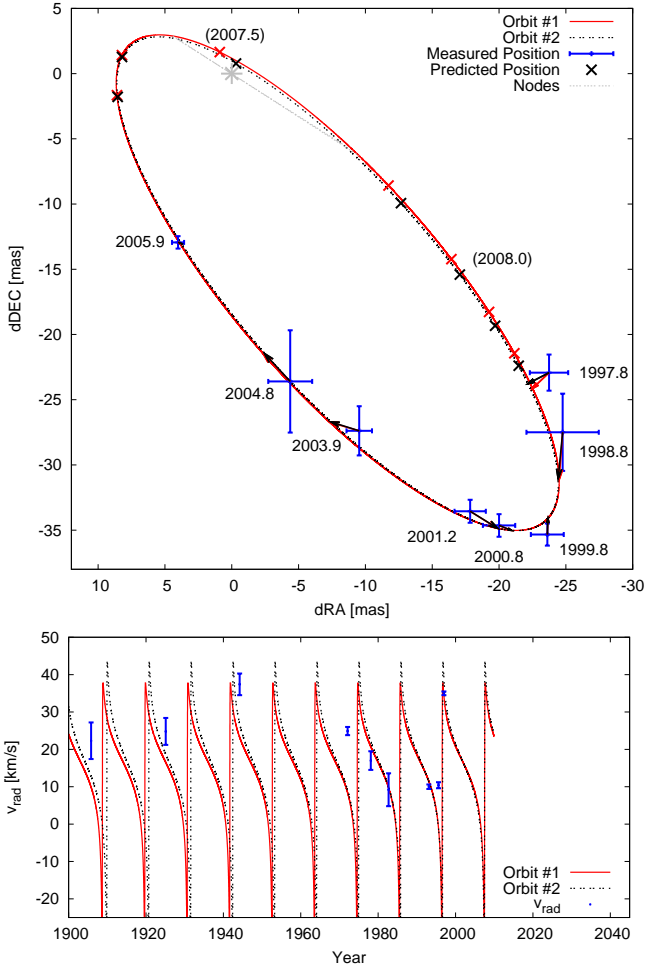}
  \caption{\textit{Top: }Two best-fit apparent orbits of the {\toric} binary
    system (solid and dashed line).  Each position measurement (see
    Table~\ref{tab:position}) is connected to its predicted position with an
    $O$--$C$ line.  Besides our orbital solutions (see
    Table~\ref{tab:orbitalelements}), the line of nodes and the predicted
    positions for the interval 2007.0 to 2008.5 (increments of 0.25~yrs) are shown. 
    North is up and east is to the left.
    \textit{Bottom: }Radial velocity variations of both components computed
    for our orbital solutions using the method presented by \citet{pou98}.
    For the mass ratio $M_{\rm C2}/M_{\rm C1}$, we assume 0.47, as obtained from
    our modeling of the flux ratio presented in Sect.~\ref{sec:toricnature}.
    When computing the radial velocity curve, the (unknown) system velocity
    $V_{0}$ represents a free-parameter corresponding to a velocity offset,
    which we set to $-5$~km\,s$^{-1}$ for this plot.
    The radial velocity measurements (points with errorbars) were taken from
    \citet{vit02}.
  }
  \label{fig:orbit}
\end{figure}

\begin{table}[t]
  \caption{Preliminary orbital solutions, dynamical parallaxes, and system
    masses for {\toric}.}
\label{tab:orbitalelements}
\centering
\begin{tabular}{rc|c|c}
  \hline\hline
                      & & Orbit \#1  & Orbit \#2\\
  \hline
  $P$ & [yrs]           & 10.98      & 10.85      \\ 
  $T$ &                 & 1996.52    & 1996.64    \\
  $e$ &                 & 0.909      & 0.925      \\
  $a$ & [mas]           & 41.3       & 45.0       \\
  $i$ & [\degr]         & 105.2      & 103.7      \\
  $\Omega$ & [\degr]    & 56.5       & 56.9       \\
  $\omega$ & [\degr]    & 65.7       & 68.2       \\
  \hline
  $\chi^2/\textnormal{dof}$ &     & 1.61       & 1.59  \\
  \hline
  $\pi_{\rm dyn}$$^{a)}$    & [mas]        & $2.304 \pm 0.066$    & $2.585 \pm 0.074$ \\
  $d_{\rm dyn}$$^{a)}$      & [pc]         & $434 \pm 12$         & $387 \pm 11$ \\
  $(M_{\rm C1}+M_{\rm C2})$$^{a)}$   & [$M_{\sun}$] & $47.8 \pm 4.2$       & $44.8 \pm 3.9$ \\
  \hline
\end{tabular}
\begin{minipage}{\linewidth}
\vspace{2mm}
\hspace{5mm}Notes~--~$^{a)}$ The errors on the dynamical parallaxes and
  corresponding distances were estimated by varying the measured binary flux
  ratio within the observational uncertainties, the assumed spectral types
  for the bolometric correction by one sub-class, the extinction by $\pm 0.2$
  magnitudes, and by using three different MLRs
  \citep[by][]{bai46,hei78,dem91}.  However, the given errors do not reflect
  the uncertainties on the orbital elements $a$ and $P$.
  Due to the presence of the multiple orbital solutions, it is currently
  not possible to quantify these errors reliably.
\end{minipage}
\end{table}

\begin{table}[t]
\caption{Ephemerides for the {\toric} orbital solutions presented in
                      Table~\ref{tab:orbitalelements}}
\label{tab:ephemerides}
\centering
\begin{tabular}{lllccccccccc}
  \hline\hline
  Epoch    &  \multicolumn{2}{c}{Orbit \#1} & & \multicolumn{2}{c}{Orbit \#2}\\
  \cline{2-3}\cline{5-6}
           &  $\Theta$  & $\rho$     & & $\Theta$   & $\rho$\\
           & [\degr]    & [mas]      & & [\degr]     & [mas]\\
  \noalign{\smallskip}
  \hline
  \noalign{\smallskip}
  2007.0 & 100.9 &  8.7 && 101.8 &  8.7\\
  2007.5 &  28.6 &  1.9 && -22.4 &  0.8\\
  2008.0 & 229.1 & 21.7 && 227.9 & 23.0\\
  2008.5 & 224.6 & 30.1 && 223.8 & 31.0\\
  2009.0 & 221.7 & 35.1 && 221.1 & 35.8\\
  2010.0 & 217.5 & 40.2 && 217.0 & 40.5\\
  2011.0 & 213.9 & 41.6 && 213.4 & 41.6\\
  2012.0 & 210.3 & 40.5 && 209.8 & 40.2\\
  2013.0 & 206.3 & 37.5 && 205.8 & 37.0\\
  2014.0 & 201.4 & 33.0 && 200.8 & 32.3\\
  2015.0 & 194.6 & 27.1 && 193.7 & 26.3\\
  2016.0 & 183.4 & 20.1 && 181.7 & 19.2\\
  2017.0 & 159.7 & 12.8 && 155.1 & 12.0\\
  \noalign{\smallskip}
  \hline
\end{tabular}
\end{table}

Our multi-epoch position measurements of the {\toric} system can be used to
derive a preliminary dynamical orbit.  To find orbital solutions, we used the
method described by \citet{doc85}.  This method generates a class of Keplerian
orbits passing through three base points. From this class of possible
solutions, those orbits are selected which best agree with the measured
positions, where we use the error bars of the individual measurements as
weight.  
In order to avoid over-weighting the orbit points which were
sampled with several measurements at similar epochs (two measurements in 
1999.7-1999.8 and three measurements in 2003.8-2003.9), we treated each of
these clusters as single measurements.

In Table~\ref{tab:orbitalelements} we give the orbital elements corresponding
to the two best orbital solutions found.  
As the $\chi^2/\textnormal{dof}$ values of the two presented orbits are
practically identical, the existing astrometric data does not allow us
to distinguish between these solutions.  These orbits and the corresponding
O--C vectors are shown in Figure~\ref{fig:orbit} (see Table~\ref{tab:position}
for a list of the O--C values).  As the ephemerids in
Table~\ref{tab:ephemerides} and also the position predictions (dots) in
Figure~\ref{fig:orbit} show, future high-accuracy long-baseline
interferometric measurements are needed to distinguish between these
orbital solutions.

Potentially, additional constraints on the {\toric} binary orbit could be
provided by radial velocity measurements, such as those published by
\citet{vit02} and in the references therein.
However, the complexity of the {\toric} spectrum -- including the line
variability corresponding to the magnetically confined wind-shock region
expected towards {\toric} -- makes both the measurement and the interpretation
of radial velocities for {\toric} very challenging. Since it is
unclear whether these velocities really correspond to the orbital
motion of the binary system or perhaps to variations in the stellar wind from
\toric, we did not include these velocity measurements as a tough constraint in
the final orbital fit, but show them together with the radial velocities
corresponding to our best-fit orbit solutions in Figure~\ref{fig:orbit}.

Both orbital solutions suggest that during periastron passage, the
physical separation between C1 and C2 decreases to $\sim1.5$~AU,
corresponding to just $\sim30$ stellar radii.
Besides the strong dynamical friction at work during such a close passage,
strong wind-wind interaction can also be expected.

It is worth mentioning that besides the presented best-fit orbital solutions, a
large number of solutions with longer orbital periods exist, which are also
fairly consistent with the astrometric measurements. However, since these orbits
have slightly higher $\chi^2/\textnormal{dof}$ values than the solutions presented
above and also correspond to physically unreasonable masses ($M_{\rm
  C1}~+~M_{\rm C2}~\lesssim~20$ or $\gtrsim 130~M_{\sun}$, assuming $d$=440~pc),
we rejected these formal solutions.

\subsection{Dynamical masses and parallaxes} \label{sec:dynmassparallax}

Kepler's third law relates the major axis $a$ and the orbital period $P$ with
the product of the system mass and the cube of the parallax;
i.e.\ $(M_{\rm C1}+M_{\rm C2}) \cdot \pi^3 = a^3/P^2$ (where $a$ and $\pi$ are
given in mas, $P$ in years, and $M$ in solar masses).

In order to separate the system mass and the parallax 
in absence of spectroscopic orbital elements, the method by \citet{bai46} can be
applied.  This method assumes that the component masses follow a
mass-luminosity relation (MLR), which, together with a bolometric
correction and extinction-corrected magnitudes, allows one to solve for the
system mass $M_{\rm C1}+M_{\rm C2}$ and the dynamical parallax $\pi_{\rm dyn}$.
When using the MLR by \citet{dem91}, the bolometric
correction for O5.5 and O9.5 stars by \citet{mar05}, and the extinction
corrected magnitudes given in Table~\ref{tab:colors}, we derive the dynamical
masses and parallaxes given in Table~\ref{tab:orbitalelements}.
When comparing the distances corresponding to the
dynamical parallaxes derived for Orbit \#1 ($d_{\rm dyn}=\pi^{-1}_{\rm dyn} =
434$~pc) and Orbit \#2 ($d_{\rm dyn}=387$~pc) with distance estimates from
the literature (e.g.\ $d=440 \pm 34$~pc from \citealt{jef07}; see also
references herein), orbit solution \#1 appears much more likely.  
The dynamical system mass corresponding to Orbit \#1 is
$47.8~M_{\sun}$, which must be scaled by a factor $(d/d_{\rm dyn})^3$ when
distances other than $d_{\rm dyn}=434$~pc are assumed.

\subsection{The orbital parameters in the context of reported periodicities}

Several studies have already reported the detection of periodicity in the
amplitude, width, or velocity of spectral lines around {\toric}.  This makes
it interesting to compare whether one of those periods can be attributed to
the presence of companion C2:

\begin{description}
\item[{\bf \boldmath $P\approx15.422 \pm 0.002$~d: \boldmath}] By far,
  the best-established periodicity towards {\toric} was detected in hydrogen 
  recombination lines and various photospheric and stellar-wind lines
  \citep{sta93,sta96, wal94, oud97}. Later, the same period was also found in
  the X-ray flux \citep{gag97} and even in modulations in the Stokes
  parameters \citep{wad06}. Although possible associations with a hypothetical
  low-mass stellar companion were initially discussed \citep{sta96}, this
  period is, in the context of the magnetic rotator model, most often
  associated with the stellar rotation period.  We can rule out that C2 is
  associated with this periodicity, as we do not see significant motion of C2
  within the seven days covered by the IOTA measurements.

\item[{\bf \boldmath 60~d$~<~P~<~2$~yrs, $P\approx120$~yrs: \boldmath}]
  \citet{vit02} fitted radial velocity variations assuming the presence of two
  companions and determined possible periods of $729.6/L$~days (with $L$ an
  integer $<13$) for the first and $\sim120$~yrs for the second
  companion.  Since our orbital solutions do not match any of these
  periods, we consider an association of {\toric}2 very unlikely. 

\item[{\bf \boldmath$~P~\gtrsim~8$~yrs: \boldmath}] \citet{sta98} reported a
  steady increase in radial velocity. \citet{don02} confirmed this trend and
  estimated that this increase might correspond to the orbital motion of a
  companion with a period between 8~yrs (for a highly eccentric orbit) and
  16~yrs (for a circular orbit). With the found period of $\sim11$~yrs, it is
  indeed very tempting to associate {\toric}2 with this potential
  spectroscopic companion. However, as noted in Sect.~\ref{sec:orbitsolution},
  the set of available spectroscopic radial velocity measurements seems rather
  inhomogeneous and fragmentary and might contain observational biases due to
  the superposed shorter-period spectroscopic line variations, as noted above.
\end{description}

\subsection{Nature of the {\toric} components} 
\label{sec:toricnature}

\begin{figure}[tbp]
  \centering
  \includegraphics[width=8.5cm]{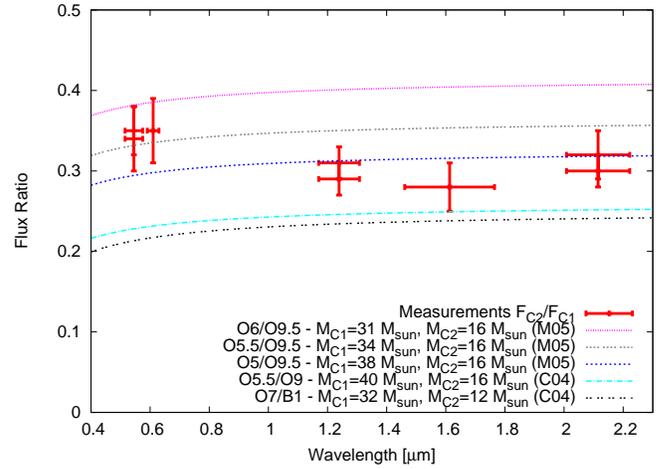}\\
  \caption{Measured intensity ratio of the {\toric} components as a function
  of wavelength (points with errorbars).  For various spectral-type
  combinations, the curves show the expected intensity ratio,
  assuming the stars contribute purely photospheric emission (black-body
  emission with luminosities and effective temperatures as given in the
  stellar evolution models from \citet[][ M05]{mar05} and
  \citet[][ C04]{cla04}).}
  \label{fig:fluxratio}
\end{figure}

\begin{figure}[tbp]
  \centering
  \includegraphics[width=8.5cm]{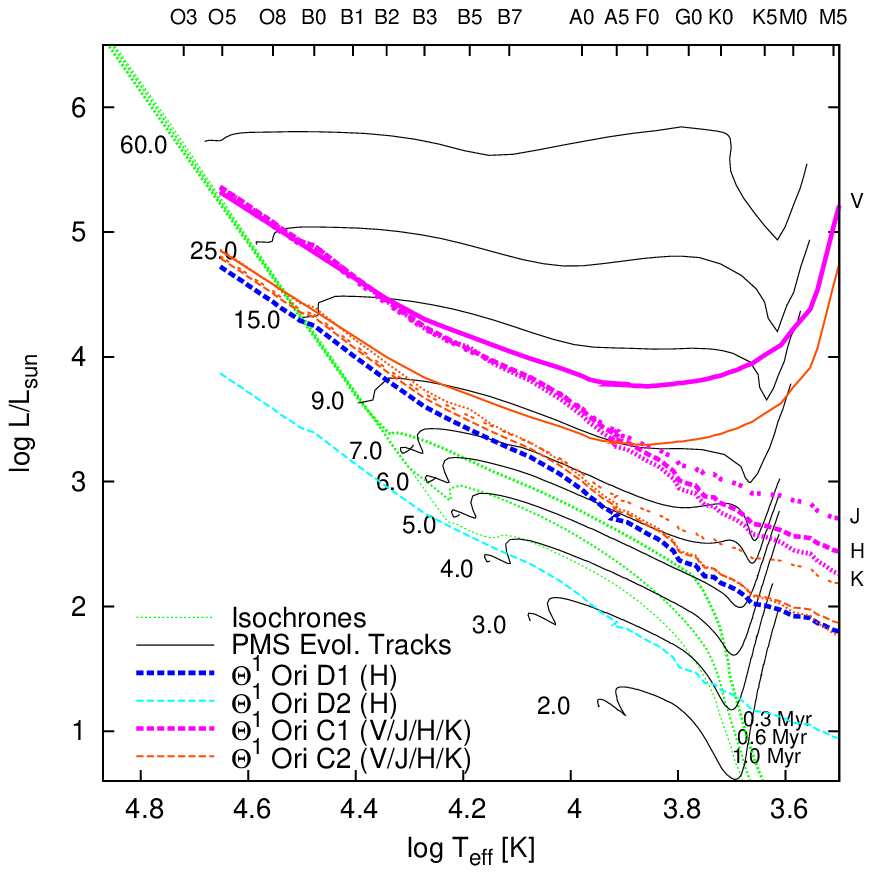}\\
  \caption{HR-diagram with PMS evolutionary tracks ($Z=0.02$; thin solid lines
  with the corresponding masses labeled aside). For masses $\leq 7
  M_{\sun}$ ($\geq 9 M_{\sun}$), the tracks from \citealt{sie00} \citep{ber96}
  were used. The three dotted lines represent the isochrones for 0.3, 0.6, and
  1.0~Myr \citep{ber96}. Using the synthetic colors and bolometric
  corrections compiled by \citet[][ including data from
  \citealt{joh66}, \citealt{bes88}, and others]{ken95} and the O star colors
  by \citet{mar05} and \citet{mar06}, we converted the measured photometry
  for {\toric}1, C2, D1, and D2 into the corresponding allowed
  locations in the HR-diagram (thin red lines). }
  \label{fig:hrd}
\end{figure}

\begin{table*}[t]
\caption{Derived dereddened magnitudes and colors for the {\toric}
  components. For the photometry of the total system, we used data
  from \citet[][ $J$=4.63, $H$=4.48, $K$=4.41]{hil98} and \citet[][
  $V$=5.12]{hil97}.  An extinction of $A_V=1.74$~\citep{hil97} was
  assumed (using the reddening law by \citealt{mat90} and, similar to
  \citealt{mat81}, a high $R_V=5.0$).}
\label{tab:colors}
\centering

\begin{tabular}{lcccc|cccccc}
  \hline\hline
                      & V    & J    & H    & K    & V--J  & V--H  & V--K  & J--H  & J--K  & H--K \\
  \hline
  {\toric}1           & 3.70 & 4.35 & 4.38 & 4.49 & -0.65 & -0.69 & -0.80 & -0.04 & -0.15 & -0.11 \\  
  {\toric}2           & 4.87 & 5.65 & 5.81 & 5.73 & -0.78 & -0.94 & -0.86 & -0.15 & -0.08 &  0.07 \\  
  \hline
\end{tabular}
\end{table*}

Most studies which can be found in literature
attributed the whole stellar flux of {\toric} to a single component and
determined a wide range of spectral types including O5.5 \citep{gag05}, O6
\citep{lev76,sim06}, O7 \citep{van88}, to O9 \citep{tru31}.  In order to
resolve this uncertainty, it might be of importance to take the
presence of {\toric}2 into account.
Besides the constraints on the dynamical masses derived in
Section~\ref{sec:orbitsolution}, additional information about the spectral
types of {\toric}1 and C2 can be derived from the flux ratio of the components
in the recorded bands.

In contrast to our earlier studies \citep{wei99,sch03a}, we can now
also include the $V$-band flux ratio measurement to constrain the spectral
types of the individual components.  The $V$-band is of particular interest,
as a relative increase of the flux ratio $F_{\rm C2}/F_{\rm C1}$ from the visual to
the near-infrared would indicate the presence of circumstellar material,
either as near-infrared excess emission or intrinsic extinction towards C2
(assuming similar effective temperatures for both components).
Our speckle measurements indicate that $F_{\rm C2}/F_{\rm C1}$ stays rather constant
from the visual to the near-infrared.  Therefore, in the following we assume
that the major contribution of {\toric}2 to the measured flux is photospheric.

In Figure~\ref{fig:fluxratio} we show the measured $F_{\rm C2}/F_{\rm C1}$ as a
function of wavelength and compare it to model curves corresponding to
various spectral-type combinations for C1 and C2.  To compute the model flux
ratios, we simulate the stellar photospheric emission as black-body emission
$B(T_{\rm eff})$ with effective temperatures $T_{\rm eff}$ and stellar radii $R$,
as predicted by stellar evolutionary models \citep{cla04,mar05}:
\begin{equation}
  \left(\frac{F_{\rm C2}}{F_{\rm C1}}\right)(\lambda) = \frac{B(T_{\rm eff}^{\rm C2}) R_{\rm C2}^2}{B(T_{\rm eff}^{\rm C1}) R_{\rm C1}^2}
\end{equation}
Under these assumptions, the companion C2 would have to be rather massive
($M_{\rm C2}/M_{\rm C1} = 0.45 \pm 0.15$) to obtain reasonable agreement with the
measured flux ratios (see Figure~\ref{fig:fluxratio}).

Using a value for $A_{V}$ from literature, the flux ratios can also be used to
estimate the photometry of the individual components
(Table~\ref{tab:colors}).  Then, the spectral type of C1 and C2 can be
determined by comparing the location of the stars in the HR-diagram with
stellar evolution models. For this, we adopt the procedure from \citet{sch03a}
and convert the derived photometry into locations in the HR-diagram using the
colors and bolometric 
corrections from \citet[][ and references therein]{ken95} and \citet{mar06}.  Assuming coevality for  
both stars, the spectral type of the individual components can be constrained
by finding the location where the curves for the various spectral bands and
the isochrone intersect.  
As can be seen in Figure~\ref{fig:hrd}, the allowed locations for C1 intersect
the Zero-Age Main Sequence\footnote{With a dynamical age of $\sim3\times10^5$~yrs, it seems
  justified that the Trapezium stars are real ZAMS stars \citep{sch03c},
  although the strong magnetic activity from {\toric} was also associated with
  a pre-main-sequence origin~\citep{don02}.} (ZAMS)
around $T_{\rm eff}=46\,000 \pm 4\,000$~K, $\log L/L_{\sun}=5.3 \pm 0.2$ (corresponding to O5) and
around $T_{\rm eff}=33\,000 \pm 2\,000$~K, $\log L/L_{\sun}=4.5 \pm 0.1$ (corresponding to O9) for C2.

We conclude that the spectral type combination, which simultaneously provides
good agreement to the measured flux ratios, the HR-diagram, and the
dynamical masses derived in Sect.~\ref{sec:orbitsolution}, is given by the
following stellar parameters (using the evolutionary models from \citealt{mar05}):\\
C1:~O5.5~($M=34.0~M_{\sun}$, $T_{\rm eff}=39\,900$~K, $\log L/L_{\sun}=5.41$)\\
C2:~O9.5~($M=15.5~M_{\sun}$, $T_{\rm eff}=31\,900$~K, $\log L/L_{\sun}=4.68$)

\subsection{Nature of the potential {\torid} companion} \label{sec:toridnature}

Although the {\torid} binary parameters presented in
Sect.~\ref{sec:t1oridmodel} must be considered preliminary, it might be
interesting to determine the spectral type of the putative components.
We apply the procedure discussed in Sect.~\ref{sec:toricnature} to
determine the photometry of the components from the measured intensity ratio
(photometry for the unresolved system from \citealt{hil98}: $H$=5.84) and
derive $H_{D1}$=5.98 and $H_{D2}$=8.12, respectively.

Searching again for the intersection between the allowed locations in the
HR-diagram with the isochrones applicable to the ONC (Figure~\ref{fig:hrd}),
the best agreement for D1 can be found with $T_{\rm eff}=31\,500 \pm
4\,000$~K, $\log L/L_{\sun}=4.25 \pm 0.1$ (corresponding to O9.5).
Accordingly, D2 might be either a B4 or B5 type star which has just reached
the ZAMS ($T_{\rm eff}=16\,000 \pm 4\,000$~K, $\log L/L_{\sun}=2.6 \pm 0.2$)
or a pre-main-sequence K0 type star ($T_{\rm eff}=5\,000 \pm 1\,000$~K, $\log
L/L_{\sun}=1.3 \pm 0.2$).

\citet{vit02b} examined radial velocity variations of {\torid} and presented
preliminary spectroscopic orbital elements for a companion with a 20.2~d
period (or twice that period, P=40.5~d).
Assuming $20~M_{\sun}$ as the system mass, these periods correspond to a major axis
of 0.05 or 0.08~AU ($\sim0.1$ or $0.2$~mas).  Since this is far below the
18~mas suggested by our binary model fit, we do not associate our potential
companion with the proposed spectroscopic companion. 

The multiplicity rate in a young stellar population such as the Trapezium
cluster is an important quantity which might allow us to draw conclusions
not only about the dynamical history of the ONC, but also
about the mechanisms controlling the star formation process.
The detection of a new companion around {\torid} further increases
the multiplicity rate for high-mass stars in the ONC.  For instance,
considering the sample of 13 Orion O- and B-type stars studied by
\citet{pre99} now yields 10 visual and 5 spectroscopic detected companions
(including one quintuple system, namely {\torib}).  This corresponds to an
average observed companion star frequency (CSF) of 1.15 companions per
primary.  Despite the fact that this value only represents a strict lower
limit due to observational incompleteness, it is already higher than
the incompleteness-corrected CSF determined by \citet{duq91} for a
distance-limited sample of solar-type field stars (0.5 companions per
primary).  \citet{koe06} have reported that the CSF for low- and
intermediate-mass stars in the ONC is about a factor of 2.3 lower than the
CSF in the \citeauthor{duq91} sample, making the differences in the CSF
between the low-, intermediate-, and high-mass star population in the ONC
highly significant.  Several studies \citep[e.g.\ ][]{pre99,bal05,
  bon05} have already interpreted this as evidence that different formation mechanisms
(e.g.\ stellar coalescence vs.\ accretion) might be at work in different
mass regimes.

\section{Conclusions}

We have presented new bispectrum speckle interferometric and infrared
long-baseline interferometric observations of the Orion Trapezium stars
{\toric} and D.  This data was used to reconstruct diffraction-limited NIR and
visual speckle images of the {\toric} binary system and, to our knowledge, the
first model-independent, long-baseline aperture-synthesis image of a young star
at infrared wavelengths.

For {\torid}, we find some indications that the system was resolved by the
IOTA interferometer.  Although the non-zero closure phase signal suggests
asymmetries in the brightness distribution (maybe indicative of a close
companion star), further observations are required to confirm this finding.

From our multi-epoch observations on {\toric} (covering the interval 1997.8 to
2005.8), we derived the relative position of the companions using
model-fitting techniques, clearly tracing orbital motion.  We presented two 
preliminary orbital solutions, of which one can be favoured due to
theoretical arguments. 
This solution implies a period of 10.98~yrs, a semi-major axis of 41.3~mas,
a total system mass of $\sim48 M_{\sun}$, and a distance of 434~pc.  Furthermore,
we find strong indications that {\toric}2 will undergo periastron passage in
mid 2007.  As the binary separation at periastron is expected to be
$\sim1$~mas, further long-baseline interferometric observations on {\toric} are
urgently needed to refine the orbital elements, the stellar masses, and
orbital parallaxes. Through comparison with stellar evolutionary
models and modeling of the measured intensity ratio, we find evidence that the
companion {\toric}2 is more massive ($M_{\rm C2}/M_{\rm C1} \approx 0.45 \pm
0.15$) than previously thought; likely of late O (O9/9.5) or early B-type
(B0).  The contribution of the companion to the total flux of {\toric} and the
interaction between both stars
might be of special importance for a deeper understanding of this intriguing
object.
Therefore, we strongly encourage observers to acquire high dispersion
spectra of the system in order to trace the expected radial velocity
variations and the wind-wind interaction of the system.

\begin{acknowledgements}
We appreciate support by the IOTA technical staff, especially M.~Lacasse and P.~Schuller.
We would like to thank the anonymous referee for helpful comments which 
improved the paper.
SK was supported for this research through a fellowship from the International
Max Planck Research School (IMPRS) for Radio and Infrared Astronomy at the
University of Bonn.
\end{acknowledgements}

\bibliographystyle{aa}
\bibliography{6965}

\end{document}